\documentclass[12pt]{article}
\usepackage{epsfig,multicol}
\usepackage[english]{babel}
\usepackage{graphicx}
\usepackage{rotating}
\usepackage{amsmath}
\usepackage{amssymb}
\usepackage{flafter}

\newcommand{\gluino}{\ensuremath{\widetilde g}}

\begin{document}

\begin{titlepage}

\begin{flushright}
SISSA 15/2004/EP 
\end{flushright}

\begin{center}

\vspace*{2cm}
{\LARGE \mbox{Gluino Annihilations and Neutralino Dark Matter}}\\
\vspace*{1.5cm}
{\bf\Large S.~Profumo, C.~E.~Yaguna}\\[0.7cm]

         {\em Scuola Internazionale Superiore di Studi Avanzati,
	Via Beirut 2-4, I-34014 Trieste, Italy
and Istituto Nazionale di Fisica Nucleare, Sezione di Trieste, 
I-34014 Trieste, Italy}\\[0.7cm] 

	{\em E-mail:} {\tt profumo@sissa.it, yaguna@sissa.it} 

\vspace*{0.8cm}

\begin{abstract}\noindent We consider supersymmetric scenarios, compatible with all cos\-mo\-lo\-gi\-cal and phenomenological requirements, where the lightest SUSY particles (LSPs) are the neutralino and a quasi degenerate gluino. We study the neutralino relic abundance, focusing on gluino (co-)annihilation effects. In case the neutralino is bino-like, in a wide mass window the relic abundance is naturally driven in the correct range for the LSP to be the main cold dark matter constituent. We show that the gluino is the {\em strongest possible coannihilating partner of a bino-like neutralino in the general MSSM}. Moreover, contrary to other coannihilation scenarios, gluino pair annihilations always dominate over coannihilation processes, even at relatively large gluino-neutralino mass splittings. Finally, we present prospects for neutralino dark matter detection in the outlined framework.

\end{abstract}  
\end{center}
\vspace{1cm}

\end{titlepage}

\section{Introduction}

Supersymmetric models with a stable lightest supersymmetric particle (LSP) have been shown to be strongly constrained by requiring that the thermal relic abundance of the LSP falls within the upper bound on the dark matter density provided by cosmological observations \cite{Spergel:2003cb}. The leading supersymmetric particle candidate for dark matter is generally considered to be the lightest neutralino \cite{Munoz:2003gx}. Demanding that the relic abundance of neutralinos falls in the correct range amounts to setting {\em upper bounds} on the lightest neutralino mass, and thus on the lowest mass scale of supersymmetric particles. In this respect, at the dawn of the LHC era, it is of the utmost importance to explore the largest possible range of low-energy realizations of supersymmetry (SUSY) compatible with a supersymmetric dark matter scenario.

The relic abundance of neutralinos is set by two features of the SUSY spectrum: first, the {\em composition of the neutralino} in terms of its gauge eigenstates (bino $\widetilde B$, wino $\widetilde W$ and Higgsino $\widetilde H$), and, second, the possible presence of {\em peculiar mass spectrum realizations} (as the approximate mass degeneracy of other SUSY particles, giving rise to {\em coannihilation effects} \cite{Griest:1990kh}, or the occurrence of resonant neutralino annihilations through $s$-channel heavy Higgs-exchange \cite{Lahanas:1999uy}). In particular, if the neutralino is wino- (or higgsino-) like, the effect of a large annihilation cross section and the presence of chargino (and of next-to-lightest neutralino) coannihilations drive the relic density to relatively low values. On the other hand, a bino-like LSP tends to have an excessive relic abundance. Higgsinos or winos  may be the main dark matter constituents only if one invokes either non-thermal production \cite{Giudice:2000ex} or modifications to the standard cosmological scenario \cite{cosmomod}, or if the lightest neutralino is heavy, killing, in this latter case, any hope of detecting SUSY at the LHC\footnote{Higgsino and wino dark matter, on the other hand, are expected to have large detection rates at direct and indirect detection experiments, though, once again, suppressed in the case of very heavy LSPs.} \cite{Edsjo:1997bg,Baer:2003wx}. As regards a bino-like LSP, as it is the case in many popular frameworks, e.g. in most of the minimal supergravity (mSUGRA) parameter space, the over-production of relic neutralinos must be compensated, outside a strongly constrained low-masses bulk region, by the mentioned coannihilation or resonance effects.

Since the discovery of coannihilation processes \cite{Griest:1990kh}, many dedicated studies analyzed a rather wide plethora of coannihilating partners: the lightest stau \cite{Ellis:1998kh} (e.g. in mSUGRA at low $m_0$ and $A_0$), the lightest stop \cite{Boehm:1999bj} (again in mSUGRA, at large $A_0$), the lightest chargino \cite{Edsjo:1997bg,Edsjo:2003us} (when the LSP is wino- or higgsino-like, chargino coannihilations are always present; this takes place, for instance, in the focus point region of mSUGRA, or in models with non-universal gaugino masses), the next-to-lightest neutralino (e.g. for higgsino-like LSP), and the lightest sneutrinos \cite{Ellis:2002iu,Profumo:2003em} and bottom squarks \cite{Profumo:2003em}.

In the present note we address the possibility that the coannihilating partner of the LSP is the {\em gluino} (GC, Gluino Coannihilation, Model). The low-energy condition for having gluino coannihilation processes is that $m_{\chi}\simeq m_3\equiv m_{\gluino}$. Being a strongly interacting particle, we expect in particular gluino-gluino annihilations to be greatly effective in reducing the LSP relic abundance. In view of what we outlined above, if the LSP is to be the {\em main dark matter component}, this scenario will be of particular interest in case the lightest neutralino is {\em bino-like}: gluino coannihilations will then provide, depending on the bino-gluino mass splitting, the required relic density suppression mechanism to obtain the correct dark matter thermal relic abundance. In particular, due to the very large gluino-gluino annihilation cross section, and to the presence of coannihilation processes which couple neutralino and gluino freeze-out, the net effect of neutralino relic abundance suppression is mainly driven by the gluino effective annihilattion cross section, even for relatively large gluino-neutralino mass splittings. We emphasise that this feature is peculiar of the Gluino Coannihilation scenario, since for any other coannihilating partner, for sufficinetly large mass splittings, the coannihilation amplitude dominates over the coannihilating partner pair annihilations. In case the lightest neutralino is higgsino or wino-like, gluino coannihilations will only be helpful at very large masses ($m_\chi\sim 1\div2$ TeV). We will show that, though featuring larger annihilation cross sections, the resulting relic abundance of winos and higgsinos which coannihilate with a quasi degenerate gluino is {\em larger} than that of coannihilating binos, due to the presence of additional effective degrees of freedom brought by charginos, and by the next-to-lightest neutralino in the case of higgsinos.

The main results we will present in the remainder of the paper are: 
\begin{enumerate}
\item The gluino is a {\em phenomenologically perfectly viable coannihilating partner}; the required mass spectrum is found for instance in superstring inspired models \cite{Chen:1996ap} or in gauge mediated SUSY breaking scenarios \cite{Raby:1997bp}.
\item Gluino coannihilations are the {\em strongest possible bino coannihilation processes} in the general MSSM.
\item The mass range for a bino-like lightest neutralino, in the presence of gluino coannihilations, extends in the {\em multi-TeV region}; in the light of point 2., the maximal bino mass in presence of a single coannihilating partner is precisely found in the gluino coannihilation region we investigate in the present note.
\end{enumerate}
In what follows we will detail and motivate these results, and we will further comment about dark matter detection perspectives for the model under scrutiny.

\section{The Model}
In the conventional mSUGRA model, gaugino masses at low energies ($m_i$) are proportional to the corresponding $\alpha_i$ obeying:
\begin{equation}
m_3:m_2:m_1\sim \alpha_3:\alpha_2:\alpha_1\sim 6:2:1~. \label{gauginos}
\end{equation}
This relation is a consequence of the {\em assumed} gaugino mass universality at high energies ($M_i=M_{1/2}$ at $M_{GUT}$), and, if valid, implies that the gluino is the heaviest gaugino. However, there are plenty of well motivated models which do not satisfy Eq.~(\ref{gauginos}) (see, for instance, ~\cite{Chattopadhyay:2003yk}). In particular, we are interested in low energy realizations of the MSSM  with gluino-neutralino quasi-degeneracy  and, therefore, with gluinos lighter than what expected from Eq.~(\ref{gauginos}). The high energy setup and some phenomenological implications of such models, as well as that of related scenarios with a gluino LSP, have been considered previously in the literature \cite{Chen:1996ap,Raby:1997bp,Baer:1998pg,Raby:1998xr}. 

One of these is the so-called \textrm{O-II} superstring inspired model, in the limit in which supersymmetry breaking is dominated by the universal size modulus. Gaugino masses, which are determined by the standard RGE coefficients and by the Green-Schwartz parameter, arise at one-loop and in the preferred range of the model typically yield either a (heavy) gluino LSP, or neutralino-gluino quasi degeneracy \cite{Chen:1996ap}. 

\begin{figure*}[!t]
\begin{center}
\includegraphics[scale=0.6]{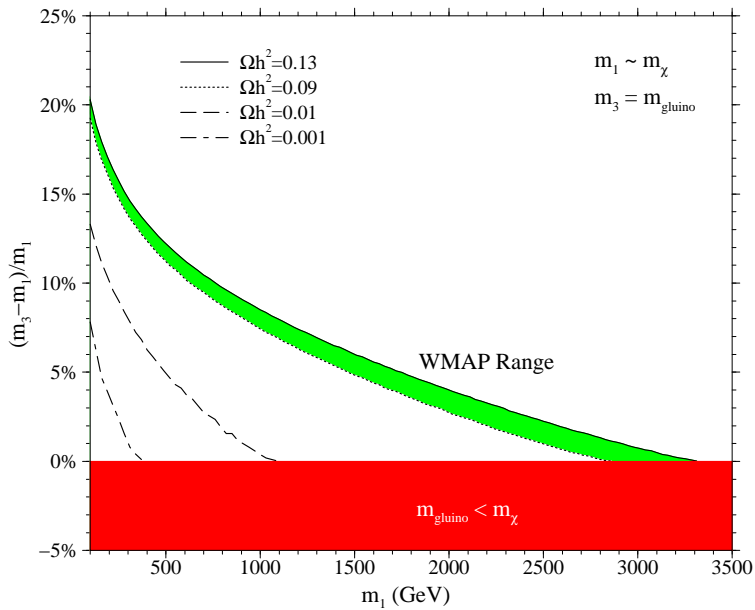}
\end{center}
\caption{\small \em The area shaded in green shows the parameter space, in the $(m_1,(m_3-m_1)/m_1)$  plane, compatible with the WMAP estimate of the cold dark matter content of the Universe. The red shaded region at negative values of $(m_3-m_1)/m_1$ is not allowed because the neutralino is no longer the LSP: a heavy gluino LSP is ruled out by anomalously heavy isotopes searches \cite{coldm}.}\label{fig:PARAMSP}
\end{figure*}

Another example emerges in the context of Gauge Meditated SUSY Breaking (GMSB). In some GUT models, as a result of the doublet-triplet splitting mechanism and due to the mixing between the Higgs and the messengers, gluino masses are suppressed ~\cite{Raby:1997bp}. Notice that in this model a smooth change in the parameter $B$, the ratio of the doublets and triplets masses, easily leads from a gluino LSP to a neutralino-gluino quasi degeneracy. 

In the present paper, we will focus on gluino (co)-annihilations processes as an effective mechanism that suppresses the neutralino relic density. It is known that bino-like neutralinos tend to produce a relic abundance well above the WMAP preferred range, whereas wino- and higgsino-like neutralinos have the opposite behavior. Hence, relic density suppression mechanisms, such as gluino coannihilations, are particularly interesting for the case of bino-like neutralinos and we will devote most of this study to that situation.  

The Gluino Coannihilation (GC) model we propose is defined as any realization of the minimal supersymmetric extension of the standard model (MSSM) satisfying the conditions 
\begin{equation}\label{eq:one}
m_\chi\ \lesssim\  m_3\ \ll\ \ m_{\rm susy},
\end{equation} 
where $m_\chi$ and $m_3$ are respectively the neutralino and the gluino masses at low energy, and $m_{\rm susy}$ stands for all other SUSY particle masses. 

Let us mention that in previous studies gluino-photino processes were considered within {\em low gaugino mass models} \cite{Farrar:1994ce,Farrar:1995pz}. There, however, the coannihilating partner was {\em not} the gluino, but rather the $R^0$ gluon-gluino hadronic bound state. Furthermore, since gaugino masses were radiatively induced in the absence of dimension-3 SUSY breaking operators, the mass range of the models was limited to the few GeV's region \cite{Farrar:1995pz}. Therefore, the whole phenomenological setup was largely different from the one we describe here.

\begin{figure*}[!t]
\begin{center}
\includegraphics[scale=0.55]{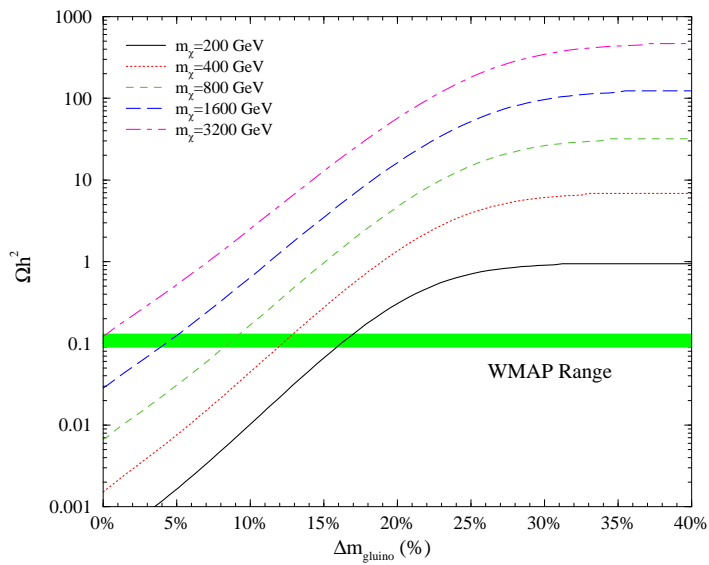}
\end{center}
\caption{\small \em The relic density as a function of the mass splitting between the neutralino and the gluino for different values of the neutralino mass. The WMAP range is shown as a green band.}\label{fig:COANN}
\end{figure*}

In what follows we will study the parameter space of the GC Model using as pa\-ra\-me\-ters $m_\chi$ and the ratio $\displaystyle\frac{m_3-m_\chi}{m_\chi}$, arbitrarily setting $m_{\rm susy}=a\cdot m_\chi$, with the numerical factor $a>1$, in order to single out the specific features of gluino coannihilations. For definiteness we fixed all the (flavor diagonal) scalar soft breaking masses $m_{\tilde s}=a\cdot m_\chi$, with $a=3$. Let us remark that any other free parameter of the MSSM, as the sign of $\mu$, $\tan\beta$, the scalar trilinear couplings $A_i$ and possible phases are largely irrelevant to the following analysis: in this respect we fixed $\mu>0$, $\tan\beta=30$ (when not otherwise specified), $A_i=0$, and any imaginary phase to zero.

The numerical study is performed through the most recent versions of the packages {\tt micrOMEGAs} \cite{Belanger:2002nx} and {\tt DarkSUSY} \cite{Gondolo:2002tz}, as respectively pertains the relic density computations and the dark matter detection rates\footnote{The current version of {\tt DarkSUSY} does not include processes with the gluino in the initial state \cite{Gondolo:2002tz}, but this does not affect the dark matter detection rates computations, for which the package is used here.}. We do not include here the perturbative cross sections for gluino-gluino and neutralino-gluino processes, which can be found elsewhere in the literature (see e.g. \cite{Baer:1998pg}). Changing the parameter $a$ slightly affects the computation of the relic density, since it varies the masses of the SUSY particles exchanged in the tree-level (co-)annihilation processes, but it leaves our analysis and our conclusions absolutely unchanged.

In fig.~\ref{fig:PARAMSP} we show, in the $\displaystyle\left(m_1,\ \frac{m_3-m_1}{m_1}\right)$ plane, the parameter space of the GC scenario for a bino-like neutralino ($m_1\approx m_\chi$). The region shaded in green corresponds to a value of the relic density compatible with the WMAP result $\Omega_{\rm CDM} h^2=0.1126^{+0.0161}_{-0.0181}$ \cite{Spergel:2003cb}. Below the green strip the relic density is over-suppressed. We show in this region isolevel curves corresponding to $\Omega h^2=0.01\,,0.001$. 

\begin{figure*}[!t]
\begin{center}
\includegraphics[scale=0.55]{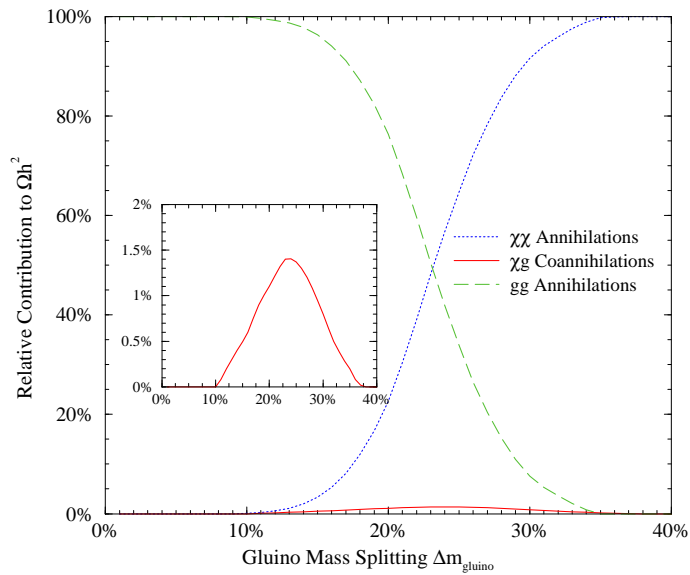}
\end{center}
\caption{\small \em Relative contributions to the relic density of $\chi\chi$ annihilations, $\chi\tilde g$ coannihilations and $\tilde g\tilde g$ annihilations as a function of the mass splitting between the gluino and a bino-like neutralino.}\label{fig:CHANN}
\end{figure*}

As expected, along the allowed region, the larger the neutralino mass, the smaller the mass splitting which ensures the needed relic density suppression. Notice that the neutralino, which is bino-like, can be as heavy as $m_\chi\sim 3.3$ TeV without entering in conflict with the constraint on the relic abundance. We recall that in mSUGRA models the upper bound on the mass of a bino-like neutralino is found to be $m_\chi\lesssim 600$ GeV \cite{Ellis:2003cw}. Let us mention that for all parameter space points we considered, direct SUSY particle searches and indirect accelerator limits on rare processes put weaker bounds than that coming from cosmology.
 
Fig.~\ref{fig:COANN} shows the relic density as a function of the mass splitting between a (bino-like) neutralino and the gluino $\Delta m_{\gluino}\equiv(m_{\gluino}-m_\chi)/m_\chi$ for different values of the neutralino mass. As  $\Delta m_{\gluino}$ increases, $\Omega_\chi h^2$ approaches its asymptotic value in the absence of coannihilations. In particular, a neutralino with a mass $m_\chi=200\,$ GeV requires a gluino with a mass splitting of about $16\%$ ($m_{\gluino} \approx 232\,$ GeV) in order to obtain a relic density within the WMAP range. If the neutralino mass is $m_\chi=1600\,$ GeV, the required splitting falls to about $5\%$ ($m_{\gluino}\approx 1680\,$ GeV). Finally, a 3 TeV neutralino needs a nearly complete gluino degeneracy in order to fulfill the upper bound on the relic abundance, as also emerging from fig.~\ref{fig:PARAMSP}.

\subsection{Gluino (Co-)annihilations}

When the gluino is quasi-degenerate with the neutralino there are three sets of processes that contribute to the evaluation of the neutralino relic density: $a)$ The usual neutralino-neutralino ($\chi\chi$) annihilations. $b)$ The neutralino-gluino ($\chi\gluino$) coannihilations. $c)$ The gluino-gluino ($\gluino\,\gluino$) annihilations. In fig.~\ref{fig:CHANN} we show the relative contribution to the effective cross section which determines $\Omega h^2$ of these three channels as a function of the gluino mass splitting $\Delta m_{\gluino}$ in the case of a {\em bino-like} lightest neutralino\footnote{The situation is similar for winos and higgsinos, though the transition from gluino annihilations to neutralino annihilations dominance in the effective cross section takes place at smaller gluino mass splitting, since the neutralino annihilation cross section is larger.}. The rest of the spectrum is taken to be decoupled ($a=3$). As seen in fig.~\ref{fig:CHANN}, the $\gluino\,\gluino$ process dominates at small mass differences, whereas the $\chi\chi$ process dominates at larger ones. The transition between these two regimes takes place at $\Delta m_{\gluino}\approx23\%$. Remarkably, the $\chi\gluino$ coannihilations play only a minor role and never contribute more than $1.5\%$, as shown in the blown up region. This fact is a very peculiar feature of gluino coannihilations. For all other possible coannihilating partners in the MSSM there is always a region, at moderate mass splittings ($\Delta m\approx 10$-$20\%$), where coannihilations (in the strict sense) are the dominant processes.

\begin{table}[!t]
\begin{center}
\begin{tabular}{ccc}
\begin{tabular}{|c|c|r|}
\hline 
 & $t \bar t$ &  $37.8\%$\\
 & $b \bar b$ &  $31.2\%$\\
 & $u \bar u$ &  $11.1\%$\\
 $\chi$ \gluino & $c \bar c$ &  $11.1\%$\\
 & $d \bar d$ &  $4.4\%$\\
 & $s \bar s$ &  $4.4\%$\\
\hline
 &  & $100\%$\\
\hline
\end{tabular}
&

\begin{tabular}{|c|c|r|}
\hline 
 & $t \bar t$ &  $48.5\%$\\
 & $u \bar u$ &  $17.2\%$\\
 & $c \bar c$ &  $17.2\%$\\
$\chi$ \gluino & $b \bar b$ &  $5.7\%$\\
  & $s \bar s$ &  $5.7\%$\\
  & $d \bar d$ &  $5.7\%$\\
\hline
 &  & $100\%$\\
\hline
\end{tabular}

&

\raisebox{0.3cm}
{\begin{tabular}{|c|c|r|}
\hline
 & $g g$ &  $59.8\%$\\
 & $u \bar u$ &  $6.7\%$\\
 & $c \bar c$ &  $6.7\%$\\
 & $t \bar t$ &  $6.7\%$\\
 \gluino\,\gluino & $b \bar b$ &  $6.7\%$\\
 & $d \bar d$ &  $6.7\%$\\
 & $s \bar s$ &  $6.7\%$\\
 \hline
 &  & $100\%$\\
\hline
\end{tabular}
}\\
 & & \\

(a)& (b) & (c)

\end{tabular}
\caption{\small \em (a) Final states for the coannihilation process $\chi \gluino$ at {\em large} $\tan\beta=50$. (b) Final states for the coannihilation process $\chi \gluino$ at {\em low} $\tan\beta=5$. (c) Final states for the annihilation process $\gluino\,\gluino$.}\label{table}
\end{center}
\end{table}  

Table ~\ref{table} shows the different final states of $\chi\gluino$ coannihilations and of $\gluino\,\gluino$ annihilations, as well as their relative importance. $\chi\gluino$ coannihilations are $\tan\beta$ dependent, and are investigated for $\tan\beta=50$ in $(a)$ and $\tan\beta=5$ in $(b)$. Notice that the $t \bar t$ channel, due to the large top  Yukawa coupling, always gives the largest contribution. The  $b\bar b$ channel, on the other hand, is very sensitive to the value of $\tan \beta$, approaching the $t\bar t$ contribution at large $\tan\beta$. As expected, the results for the first and second generations are identical. Let us stress that, in view of the gluino-gluino dominance shown in fig.~\ref{fig:CHANN}, the inclusion of quark Yukawa couplings is largely irrelevant in the present scenario. 

Since $\gluino\,\gluino$ annihilations are driven by strong interactions, they do not depend on $\tan\beta$. In $(c)$, the possible final states for the $\gluino\,\gluino$ annihilations are shown. Notice that the purely gluonic $g$-$g$ final state gives the lion's share of the effective annihilation cross section. The other final states are quark-antiquark pairs, and all of them give the same contribution.

\begin{figure*}[!t]
\begin{center}
\includegraphics[scale=0.6]{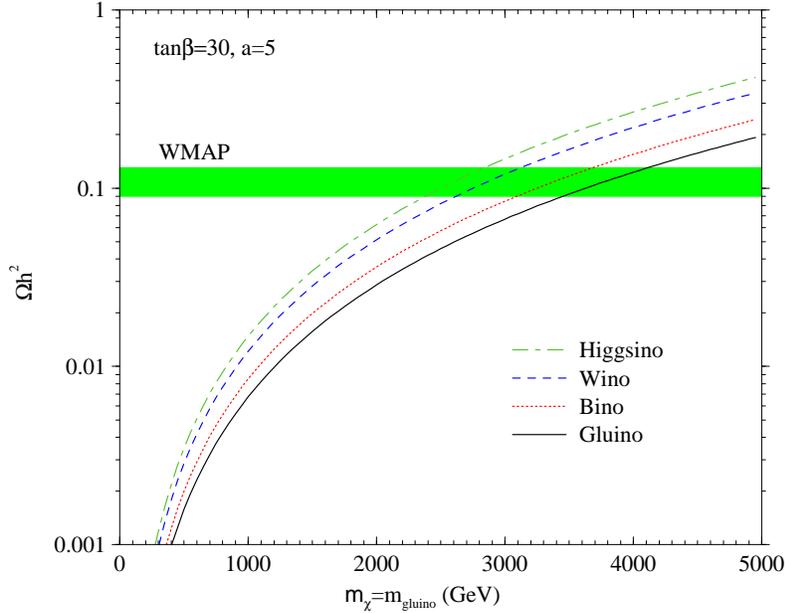}
\end{center}
\caption{\small \em The relic abundance of higgsinos, winos and binos purely degenerate in mass with the gluino. The rest of the spectrum is set to be 5 times larger than the lightest neutralino mass, and $\tan\beta=30$. We also depict the relic density of a gluino LSP, and the cold dark matter range favored by WMAP. The relic abundance of the neutralino is driven by that of the gluino, modulo an overall factor which depends on the effective degrees of freedom carried by the neutralino (see the text).}\label{fig:NEUT}
\end{figure*}

The fact that the annihilation cross section of a gluino is by far larger than that of a neutralino holds true not only if the neutralino is bino-like, but also if it is wino or higgsino-like. In this respect, we now turn to the comparison of the relic abundance of higgsinos, winos and binos which coannihilate with gluinos. We focus for clarity on the fully degenerate mass case ($m_{\chi}=m_{\gluino}$). We plot in fig.~\ref{fig:NEUT} the relic abundances for the cases of bino, wino and higgsino-gluino coannihilations. We also plot the relic density of a gluino LSP, $\Omega_{\gluino}h^2$. All other relevant SUSY masses are set to 5 times the LSP mass (this maximizes the gluino cross section, suppressing the negatively interfering $t$ and $u$ channel squark exchanges), and $\tan\beta$ is set to 30, though, clearly, the gluino cross section does not depend on it. 

Due to the gluino dominance in the effective cross section, the relic abundance of binos turns out to be the most suppressed one, as shown in fig.~\ref{fig:NEUT}. This depends on a suppression factor originating from the effective degrees of freedom which enter in the number density computation, and which, due to the neutralino and chargino mass matrix structure, depend, in their turn, on the dominant gauge component of the lightest neutralino. We emphasise that the following discussion relies on the results found in the previous section, namely on the dominance of gluino annihilation processes over coannihilations in a wide range of mass splittings, and on the presence of inter-conversion processes between the two species, which is mandatory to enable gluino annihilations to drive the neutralino relic abundance to low values affecting the neutralino freeze-out effective (co-)annihilation cross section.

The computation of the degrees of freedom suppression factor goes like this: the LSP relic density scales as the inverse of the thermally averaged effective (co-)annihilation cross section
\begin{equation}
\Omega_\chi h^2\propto \langle\sigma_{\rm eff}v\rangle^{-1}.
\end{equation}
In its turn,
\begin{equation}
\langle\sigma_{\rm eff}v\rangle(T)=\frac{A(T)}{n^2_{\rm eq}(T)},
\end{equation}
where $A$ is the annihilation rate per unit volume at a given temperature, and $n_{\rm eq}$ is the equilibrium number density, which, to a very good approximation, follows a Maxwell-Boltzmann distribution \cite{Edsjo:2003us}. The annihilation rate scales with additional degrees of freedom, in presence of coannihilation processes, as
\begin{equation}
A\rightarrow\sum_{ij}A_{ij}\frac{g_i\ g_j}{g_1^2}
\end{equation}
where the sum is extended over all annihilation and coannihilation channels, $g_i$ and $g_j$ are the degrees of freedom of the given (co-)annihilating partner, and $g_1$ are the LSP degrees of freedom. Comparing the pure gluino with the neutralino-gluino coannihilation case, since, as shown above,
\begin{equation}
A_{\widetilde g\widetilde g}\gg A_{\widetilde g \chi},\ A_{\chi \chi}
\end{equation}
the annihilation rate gives a first factor
\begin{equation}
\frac{A_{\chi-\gluino}}{A_{\gluino}}=\left(\frac{g_{\widetilde g}}{g_{\chi}}\right)^2.
\end{equation}
Further, the equilibrium number density reads
\begin{equation}
n_{\rm eq}=\frac{T}{2\pi^2}\sum_i g_i m_i^2K_2\left(\frac{m_i}{T}\right),
\end{equation}
where $K_2$ is the modified Bessel function of the second kind of order 2; since $m_{\chi}\simeq m_{\widetilde g}$, $n_{\rm eq}$ gives a second factor
\begin{equation}
\frac{n^{\chi-\gluino}_{\rm eq}}{n^{\gluino-\gluino}_{\rm eq}}=\frac{g_{\chi}}{g_{\chi}+g_{\widetilde g}}.
\end{equation}
Combining both contributions, we obtain the resulting neutralino relic density in terms of that of the gluino $\Omega_{\gluino} h^2$ (this result holds actually in the generic case of a quasi degenerate coannihilating partner whose annihilation cross section is much larger than that of the LSP):
\begin{equation}
\Omega_\chi h^2\ =\ \Omega_{\widetilde g} h^2 \left(\frac{g_{\widetilde g}+g_\chi}{g_{\widetilde g}}\right)^2
\end{equation}
In the case of the bino, since $g_{\widetilde g}=16$ and $g_\chi=2$ one gets a net increase factor equal to 1.27. 

The stated result is easily generalized to the case of other coannihilating partners $\widetilde P_i$ besides the gluino, again featuring an annihilation cross section much smaller than that of the gluino, and explains why further coannihilating partners actually {\em rise}, in this case, the final relic density:
\begin{equation}
\Omega_\chi h^2\ =\ \Omega_{\widetilde g} h^2 \left(\frac{g_{\widetilde g}+g_\chi+\sum_i g_{\widetilde P_i}}{g_{\widetilde g}}\right)^2.
\end{equation}
For instance, in the case of the higgsino one has 6 additional degrees of freedom from the next-to-lightest neutralino and from the lightest chargino, while in that of the wino there are 4 further chargino degrees of freedom. This translates into a relic density which is respectively 2.25 and 1.89 larger than that of a pure gluino. Remarkably, the numerical results nicely agree with the stated predictions (see fig.~\ref{fig:NEUT}).

We emphasize that in the present computations we neglected non-perturbative effects in the gluino-gluino scattering cross section \cite{Starkman:1990nj,Raby:1997bp,Baer:1998pg}: the evaluation of the effects of multiple gluon exchanges between interacting gluinos has in fact been shown to be highly model-dependent \cite{Baer:1998pg}. We must however warn the reader that the mentioned non-perturbative effects could {\em enhance} the gluino annihilation cross section by even orders of magnitude, and that therefore the relic density may be {\em much smaller} than what we show. In this respect, our results must be effectively regarded as conservative {\em upper bounds} on the final gluino relic density (and therefore on the coannihilating neutralino relic abundance as well). The same applies for the comparison of the efficiency of coannihilation effects we carry out in next section, as well as for the determination of an upper bound on the neutralino mass: when taken into account, non-perturbative contributions may considerably enlarge the cosmologically allowed mass ranges.

\begin{figure*}[!t]
\begin{center}
\includegraphics[scale=0.6]{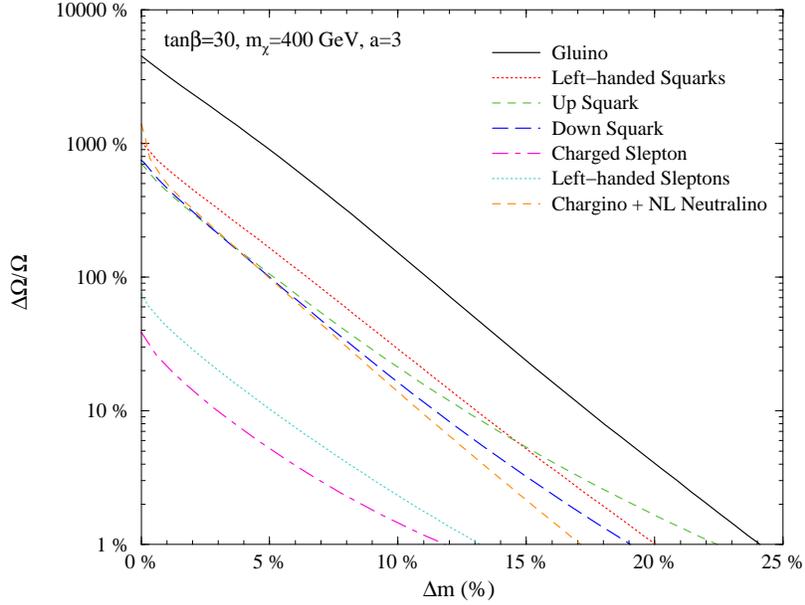}
\end{center}
\caption{\small \em The relative importance of coannihilation processes for the possible superpartners of a bino LSP in the general MSSM. $\Delta \Omega$ is the difference between the relic density computed without the coannihilation effects and the relic density computed fully including coannihilations. $\Delta m$ is the mass difference between the neutralino and the corresponding coannihilating particle. In this figure $m_\chi=400\,$ GeV, $\tan\beta=30$ and $a=m_{\rm susy}/m_{\chi}=3$.}\label{fig:COMPARE}
\end{figure*}

\subsection{Comparing the Efficiency of Coannihilations in the MSSM}

In this section we show that the gluino is the {\em strongest possible coannihilating partner of a bino-like neutralino within the MSSM}. The choice of a bino is motivated on the one hand by the fact that only in this case there are {\em no automatic chargino coannihilations}; on the other hand binos {\em require} relic density suppression mechanisms in order to produce a sufficiently small thermal relic abundance, and constitute therefore the physically more interesting case. 

In order to single out the coannihilation effects, we define $\Delta\Omega$ as the difference between the relic density obtained without taking into account coannihilation processes and the overall relic density ($\Delta\Omega=\Omega_{without\,\,coan.}-\Omega$) \cite{Edsjo:2003us}. We show in fig.~\ref{fig:COMPARE}  a plot of $\Delta \Omega/\Omega$ as a function of $\Delta m$ for all possible coannihilating particles in the MSSM\footnote{Notice that the bino purity of a bino-like lightest neutralino may be very close to 1 even if $m_1\simeq m_2$ (but {\em not} if $m_1\simeq \mu$!); in this case, it is necessary to take into account {\em simultaneously } coannihilation effects with the next-to-lightest neutralino {\em and} with the lightest chargino.}: $\gluino$, $\tilde u_R$, $\tilde d_R$, $\tilde Q_L$, $\tilde e_R$, $\tilde E_L$ and $\chi^{\pm,0}$. Again with the purpose of focusing on particular coannihilating channels, we fixed all masses of non-coannihilating super-partners to be three times the neutralino mass ($a=3$)\footnote{In this respect, our results cannot be directly compared to other assessments of coannihilation effects in particular frameworks, for instance minimal supergravity \cite{Edsjo:2003us,Ellis:1998kh}.}. The figure shows as expected that coannihilation effects are Boltzmann-suppressed by a factor which scales as $\sim {\rm e}^{-\Delta m/T}$. It is clearly seen in the plot that  gluino coannihilations are always the most effective ones, even neglecting non-perturbative contributions. For instance, if $\Delta m=5\%$, neglecting coannihilations would amount to an error in the computation of the relic density of about $10\%$ for sleptons, $100\%$ for squarks and $\chi^{\pm,0}$, and $1000\%$ for the gluino. The largest possible mass of a bino-like LSP in presence of single-partner coannihilations is therefore set, in the general MSSM, by the gluino-gluino effective annihilation cross section\footnote{The possible occurrence of an $s$-channel resonance, moreover, would not affect our conclusions. We checked in fact that in the so called ``A-pole funnel'' the bino (i.e. an LSP with a bino purity larger than 98\%) mass upper bound is always lower than what is found in the gluino coannihilation region, for any SUSY spectrum and any value of $\tan\beta$. The only caveat may be provided by unnatural multiple sfermions and/or gaugino coannihilations.}.

\subsection{Dark Matter Detection and Accelerator Searches}
\begin{figure*}[!t]
\begin{center}
\includegraphics[scale=0.49]{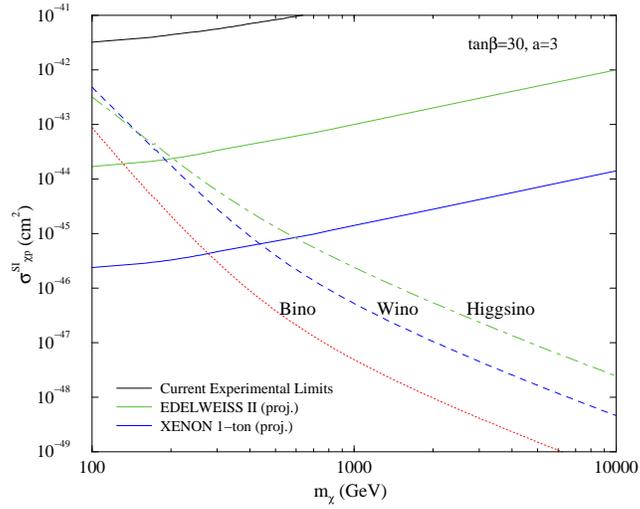}
\end{center}
\caption{\small \em The spin-independent neutralino-proton scattering cross section as a function of the neutralino mass in the purely degenerate case $m_{\chi}=m_{\gluino}$ for a bino, wino and higgsino LSP. For comparison, we also show the expected sensitivity of present and planned experiments, assuming a standard iso-thermal sphere dark matter halo profile \cite{ref:dmsens}.}\label{fig:DET}
\end{figure*}

The detection of dark matter through direct scattering on nucleons or through the detection of neutralino annihilation products from the Sun, the center of the Earth or from the center of the Galactic Halo is a rich and rapidly evolving field of research \cite{Munoz:2003gx}. Analyzing models which could provide the correct dark matter content, as it is the present case, forces therefore one to draw some conclusions on dark matter detection perspectives.

The GC scenario does not provide any significant enhancement neither in direct nor in indirect detection, since a light gluino does not affect the relevant interaction cross sections. Henceforth, our results mainly coincide with that of a SUSY model with a purely bino, wino or higgsino-like LSP and a heavy SUSY particle spectrum, with the additional possibility of having a heavy LSP, even in the multi-TeV region, thanks to gluino coannihilations. Since both direct and indirect detection rates are typically suppressed with growing LSP masses (unless peculiar cancellation mechanisms apply \cite{Ellis:2000ds}), we expect that dark matter detection in the GC scenario will not offer particularly rich perspectives.

We plot in fig.~\ref{fig:DET} the neutralino-proton spin-independent cross section for the purely degenerate case $m_{\chi}=m_{\gluino}$ in the three cases of bino, wino and higgsino-like LSP. Since the gluino does not enter in the game, the possible gluino mass splitting would not affect our results, and the only relevant physical variable is the neutralino mass. The resulting scattering cross section is compared in the figure against the current and planned experimental sensitivities, computed with a standard iso-thermal profile for the dark matter halo \cite{ref:dmsens}. Only future experiments will be able to probe the low-mass range of GC models, whose scattering cross section lies orders of magnitude below the current experimental limits. As expected, a larger higgsino or wino content yields a larger scattering cross section, as the strongest neutralino-nucleon interaction channels are those with a (light and heavy CP-even) Higgs exchange .

As regards indirect dark matter detection, the gluino mass degeneracy is not expected to yield any enhancement in the detection perspectives. In fact, we checked that in one of the most promising indirect channels, the detection of neutrinos from the decay of muons produced in neutralino annihilations captured in the center of the Sun or of the Earth, the flux typically lies, even in the low neutralino mass range, well below the planned sensitivity of future neutrino telescopes \cite{ref:antares}.

Concerning the discovery prospects of a light, coannihilating, gluino at accelerators, early detailed analysis may be found e.g. in \cite{Baer:1998pg,Raby:1998xr}. In particular, the case of wino LSP has been discussed in Ref.~\cite{Raby:1998xr}. In case the LSP is a bino, the only open channel for gluino detection would be into jets plus missing transverse energy. Particle detection at future hadronic accelerators would then be more problematic than in standard scenarios with universal gaugino masses. The LHC reach for generic GC models would then strongly depend on the specific experimental cuts adopted. A detailed assessment of the LHC SUSY discovery potential for the outlined framework, though of great interest, would however lie beyond the scope of the present analysis. 

\section{Conclusions}

In this paper we studied low energy realizations of the MSSM featuring a lightest neutralino and a quasi degenerate gluino. These models have been shown to be cos\-mo\-lo\-gi\-cal\-ly and phenomenologically viable. We highlighted the importance and the implications of gluino (co-)annihilation processes with a bino LSP, and outlined the physically relevant parameter space. We found that the LSP mass range allowed by cosmological bounds widens into the multi-TeV region, and that gluino-gluino annihilations always over-dominate gluino-bino coannihilations, contrary to any other coannihilation process. We also showed that, contrary to what naively expected, the relic density of winos and higgsinos coannihilating with the gluino is larger than that of binos. Moreover, the gluino is found to be the strongest possible coannihilating partner of a bino-like lightest neutralino in the MSSM. Finally, we briefly discussed the relevant detection perspectives for the outlined scenario at dark matter search experiments and at accelerators.



\end{document}